\pdfoutput=1
\documentclass[11pt]{article}
\usepackage[]{acl}

\usepackage{microtype}

\usepackage[T1]{fontenc}

\usepackage[hang,flushmargin]{footmisc}

\usepackage{url}            % simple URL typesetting
\usepackage{booktabs}       % professional-quality tables
\usepackage{amsfonts}       % blackboard math symbols
\usepackage{nicefrac}       % compact symbols for 1/2, etc.
\usepackage{microtype}      % microtypography
\usepackage{amsmath}
\usepackage{enumitem}
\usepackage{graphicx}       % resizebox
\usepackage{caption}
\usepackage{subcaption}
\usepackage{threeparttable}
\usepackage{color, colortbl}
\usepackage{threeparttable,booktabs}
\usepackage{multirow}
\usepackage{balance}

\usepackage{xspace,mfirstuc,tabulary}
\usepackage{tcolorbox}

\newcommand\ignore[1]{}

\begin{document}

\title{Toward Automatic Relevance Judgment using Vision--Language Models for Image--Text Retrieval Evaluation}

\author{
Jheng-Hong Yang, Jimmy Lin \\[1ex]
David R. Cheriton School of Computer Science\\
University of Waterloo, Canada\\[1ex]
\texttt{\{jheng-hong.yang, jimmylin\}@uwaterloo.ca}\\
}

\pagestyle{empty}

%% This command processes the author and affiliation and title
%% information and builds the first part of the formatted document.
\maketitle

\begin{abstract}
Vision--Language Models (VLMs) have demonstrated success across diverse applications, yet their potential to assist in relevance judgments remains uncertain. 
This paper assesses the relevance estimation capabilities of VLMs, including CLIP, LLaVA, and GPT-4V, within a large-scale \textit{ad hoc} retrieval task tailored for multimedia content creation in a zero-shot fashion.
Preliminary experiments reveal the following:
(1) Both LLaVA and GPT-4V, encompassing open-source and closed-source visual-instruction-tuned Large Language Models (LLMs), achieve notable Kendall's $\tau \sim 0.4$ when compared to human relevance judgments, surpassing the CLIPScore metric.
(2) While CLIPScore is strongly preferred, LLMs are less biased towards CLIP-based retrieval systems.
(3) GPT-4V's score distribution aligns more closely with human judgments than other models, achieving a Cohen's $\kappa$ value of around 0.08, which outperforms CLIPScore at approximately -0.096.
These findings underscore the potential of LLM-powered VLMs in enhancing relevance judgments.
\end{abstract}

\section{Introduction}
Cranfield-style test collections, consisting of a document corpus, a set of queries, and manually assessed relevance judgments, have long served as the foundation of information retrieval research~\cite{cleverdon1960aslib}. 
However, evaluating every document for every query in a substantial corpus often proves cost-prohibitive. 
To tackle this challenge, a subset of documents is selected for assessment through a pooling process. 
While this method is cost-effective compared to user studies, it has limitations due to its simplifications and struggles to adapt to complex search scenarios and large document collections.

In this study, we explore the adaptability of model-based relevance judgments for image--text retrieval evaluation. 
Leveraging model-based retrieval judgments presents an appealing option. 
Not only does it provide valuable insights before undertaking the laborious processes of document curation, query creation, and costly annotation, but it also has the potential to extend and scale up to complex search scenarios and large document collections.
To explore opportunities and meet the demands for large-scale, fine-grained, and long-form text enrichment scenarios in image-text retrieval evaluation \cite{schneider2021towards,kreiss-etal-2022-context,singh-etal-2023-enhancing,yang2023atomic}, our objective is to extend the human-machine collaborative framework proposed by \citet{faggioli2023perspectives} to the context of image-text retrieval evaluation, alongside widely adopted model-based image-text evaluation metrics \cite{hessel-etal-2021-clipscore,park2021benchmark,kim2022mutual,ruiz2023dreambooth,kreiss2023contextref}.

Our primary focus is on a fully automatic evaluation paradigm, where we harness the capabilities of Vision--Language Models (VLMs), including CLIP~\cite{radford2021learning}, as well as visual instruction-tuned Large Language Models (LLMs) like LLaVA~\cite{liu2023llava,liu2023improved} and GPT-4V~\cite{gpt4v}.
To evaluate this approach, we conducted a pilot study using the TREC-AToMiC 2023 test collection, which is designed for multimedia content creation~\cite{yang2023atomic}, based on our instruction prompt template for VLMs (cf. Table~\ref{tab:prompt} and Section~\ref{sec:model}).

We observe that model-based relevance judgments generated by visual instruction-tuned LLMs outperform the widely adopted CLIPScore~\cite{hessel-etal-2021-clipscore} in terms of ranking correlations and agreements when compared to human annotations.
While this discovery holds promise, we also uncover the potential evaluation bias when using model-based relevance judgments.
Our analysis reveals a bias in favor of CLIP-based retrieval systems in the rankings when employing model-based relevance judgments, resulting in higher overall effectiveness assessments for these systems.
In summary, our contributions can be distilled as follows:
\begin{itemize}[leftmargin=*]
    \item We demonstrate and explore the feasibility of incorporating VLMs for fully automatic image--text retrieval evaluation.
    \item We shed light on the evaluation bias when utilizing model-based relevance judgments.
\end{itemize}

\begin{table}[t]
\centering
\caption{Prompt template for relevance estimation. The VLMs are expected to take text $q$ and image $d$ independently. The prompts are only applied to the textual input $q$, while the VLMs process the pixel values of image $d$ directly.}
\begin{minipage}{\columnwidth}
\centering
\begin{tcolorbox} 
\centering
\footnotesize
\begin{tabular}{p{\columnwidth}c}
\textbf{Text Input:} & \hspace{-3.2cm} \textbf{Image Input:} \\
& \hspace{-3.3cm} \multirow{4}{*}{ \includegraphics[height=1.5cm]{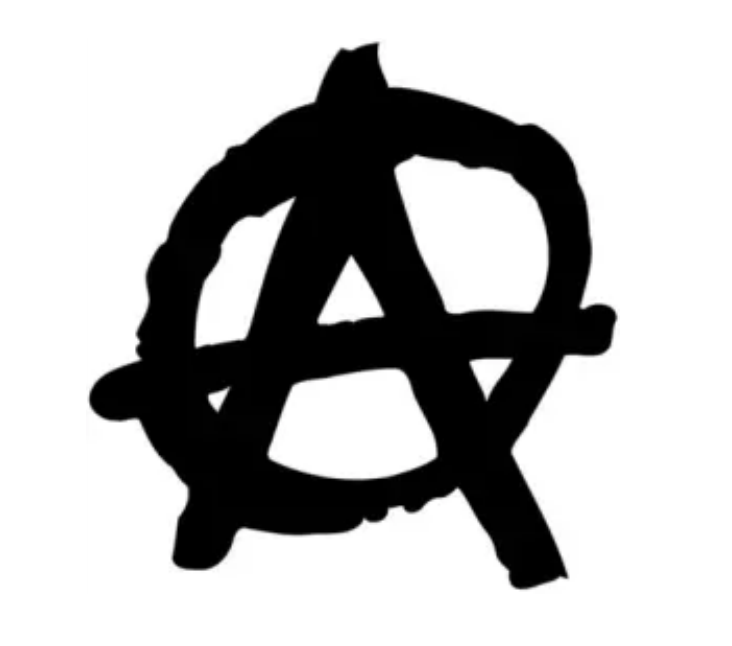}} \\
\textbf{Context:} & \\
Page Title: \texttt{<page\_title>} &\\
Page Context: \texttt{<page\_context>} &\\
Section Title: \texttt{<section\_title>} &\\
Section Context: \texttt{<section\_context>} &\\
\textbf{Relevance Instruction:} & \\
Think carefully about which images best illustrate the SECTION subject matter. Given the text and the image please answer the following questions given the criteria listed as follows: & \\
* Images must be significant and relevant in the topic's context, not primarily decorative. They are often an important illustrative aid to understanding. & \\
* Images should look like what they are meant to illustrate, whether or not they are provably authentic. & \\
* Textual information should almost always be entered as text rather than as an image. & \\
\textbf{Output Instruction:} & \\
Rate the image's overall relevance (integer, scale: 1-100) in terms of matching the text. \\
Output format: "Relevance: <score>" \\
\end{tabular}
\end{tcolorbox}
\label{tab:prompt}
\end{minipage}\vspace{-0.5cm}
\end{table}

\section{Related Work}

\paragraph{Evaluation Metrics for Image--Text Relevance.}
Nowadays, model-based evaluation metrics are widely utilized in various vision--language tasks, including image captioning~\cite{hessel-etal-2021-clipscore,chan-etal-2023-clair-evaluating} and text-to-image synthesis~\cite{park2021benchmark,Hu2023TIFAAA}.
Among model-based approaches, CLIP-based methods~\cite{park2021benchmark,kim2022mutual,Chan2023IC3IC,ruiz2023dreambooth,kreiss2023contextref}, such as CLIPScore~\cite{hessel-etal-2021-clipscore}, are particularly prevalent.
However, while these metrics are capable of measuring coarse text-image similarity, they may fall short in capturing fine-grained image--text correspondence~\cite{kreiss-etal-2022-context,yuksekgonul2023when}.
Recent research has highlighted the effectiveness of enhancing model-based evaluation metrics by leveraging LLMs to harness their reasoning capabilities~\cite{chan-etal-2023-clair-evaluating,lu2023llmscore,Betti2023LetsVM}.
There exists significant potential for incorporating LLMs into model-based approaches, as LLM outputs are not limited to mere scores but can also provide free-form texts, e.g., reasons, for further analysis and many downstream tasks~\cite{zeng2023socratic}.

\paragraph{Model-based Relevance Judgments.}
Traditionally, relevance judgments in retrieval tasks have adhered to the Cranfield evaluation paradigm due to its cost-effectiveness, reproducibility, and reliability when compared to conducting user studies.
However, this approach often relies on simplified assumptions and encounters scalability challenges.
Researchers have recently explored model-based automatic relevance estimation as a promising alternative. 
This approach aims to optimize human-machine collaboration to obtain ideal relevance judgments. 
Notably, studies of~\citet{dietz2020humans} and~\citet{faggioli2023perspectives} have revealed high rank correlations between model-based and human-based judgments. 
Additionally,~\citet{MacAvaney2023OneShotLF} have delved into the task of filling gaps in relevance judgments using model-based annotations.

\section{Methodology}
In this study, we investigate techniques for estimating image-text relevance scores, denoted as $\mathcal{F}(q, d) \in \mathbb{R}$, where $q$ represents the text (query) and $d$ represents the image (document).
Our primary focus is on utilizing VLMs to generate relevance scores, akin to empirical values annotated by human assessors denoted as $\hat{\mathcal{F}}(q, d)$.
The main objective is to assess the proximity between model-based $\mathcal{F}$ and human-based $\hat{\mathcal{F}}$ in image--text retrieval evaluation.
We begin with a discussion of the setting for human-based annotations, followed by the process for generating model-based annotations.

\subsection{Human-based Annotations}
Our primary focus revolves around a critical aspect of multimedia content creation, specifically, the image suggestion task, an \textit{ad hoc} image retrieval task as part of the AToMiC track in the TREC conference 2023 (TREC-AToMiC 2023).\footnote{\url{https://trec-atomic.github.io/trec-2023-guidelines}}
The image suggestion task aims to identify relevant images from a predefined collection, given a specific section of an article. 
Its overarching goal is to enrich textual content by selecting images that aid readers in better comprehending the material.

Relevance scores for this task are meticulously annotated by NIST assessors, adhering to the TREC-style top-$k$ pooling relevance annotation process. 
A total of sixteen valid participant runs, generated by diverse image--text retrieval systems, are considered, encompassing (CLIP-based) dense retrievers, learned sparse retrievers, caption-based retrievers, hybrid systems, and multi-stage retrieval systems. 
The pooling depth is set to 25 for eight baseline systems and 30 for the remaining participant runs. 

NIST assessors classify candidate results into three graded relevance levels to capture nuances in suitability, guided by the content of the test query. 
The test query comprises textual elements such as the section title, section context description, page title, and page context description. 
Assessors base their relevance judgments on the following criteria:
\begin{itemize}[leftmargin=*]
    \item \textbf{0 (Non-relevant)}: Candidates deemed irrelevant.
    \item \textbf{1 (Related)}: Candidates that are related but not relevant to the section context are categorized as related. They contain pertinent information but do not align with the section's context.
    \item \textbf{2 (Relevant)}: These candidates are considered relevant to the section context and effectively illustrate it.
\end{itemize}

\subsection{Model-based Annotations}\label{sec:model}
For automatic relevance estimation, we employ pretrained VLMs as our relevance estimator, denoted as $\mathcal{F}(q, d \ \lvert \mathcal{P})$.
Our relevance estimator produces relevance scores given a pair of $q$ and $d$, which is conditioned on $\mathcal{P}$, where $\mathcal{P}$ represents the prompt template we used to instruct the models. 
Prompt engineering is a commonly adopted technique for enhancing or guiding VLMs and LLMs in various  tasks~\cite{brown2020language,radford2021learning}.
It's important to note that our current focus is on pointwise estimation, leaving more advanced ranking methods (such as pairwise or listwise) that consider multiple $q$ and $d$ for future exploration~\cite{sun-etal-2023-chatgpt,qin2023large}.

\paragraph{Prompt Template Design}
In line with our approach to relevance score annotation, we have created a prompt template designed to guide models in generating relevance scores.
The prompt template, presented in Table~\ref{tab:prompt}, has been constructed based on our heuristics and is not an exhaustive search of all possible templates.
Pretrained VLMs are expected to take both $q$ and $d$ to produce a relevance score following the instructions defined in the prompt template $\mathcal{P}$. 
We anticipate that VLMs will independently process textual and visual information, and our prompt template is only applied to textual inputs.
Our template comprises three essential components:
\begin{itemize}[leftmargin=*]
    \item Context: This section processes the textual information from $q$.\footnote{For VLMs with limited context windows, e.g., CLIP, we only take the texts in the context part and ignore all the rest instructions.}
    \item Relevance Instruction: It incorporates task-specific information designed to provide VLMs with an understanding of the task.
    \item Output Instruction: This component offers instructions concerning the expected output, e.g., output types and format.
\end{itemize}

\paragraph{From Scores to Relevance Judgments.}
We utilize parsing scripts to process the relevance scores generated by the models and convert them into relevance judgments.\footnote{For CLIP, relevance scores are computed using text and image embeddings directly.}
Considering potential score variations across different models, we apply an additional heuristic rule to map these scores into graded relevance levels: 0 (non-relevant), 1 (related), and 2 (relevant).
Specifically, scores falling below the median value are categorized as 0; scores within the 50-75th quantile range are designated as 1; and scores exceeding the 75th quantile are assigned a relevance level of 2.

\begin{table*}[ht!]
\centering
% \small
\renewcommand{\arraystretch}{0.9}
\caption{Ranking correlation and judgment agreement analysis. Correlations are reported in terms of Kendall's $\tau$, Spearman's $\rho_{s}$, and Pearson's $\rho_{p}$, whereas judgment agreement is reported in terms of Cohen's $\kappa$ when comparing to NIST qrels.}
\label{tab:main}
\resizebox{.95\linewidth}{!}{
\begin{tabular}{ll|ccc|ccc|c}
\toprule
\multicolumn{2}{c}{} & \multicolumn{3}{c}{NDCG@10} & \multicolumn{3}{c}{MAP} & Agreement \\
\cmidrule(lr){3-5} \cmidrule(lr){6-8} \cmidrule(lr){9-9}
Model & Version & $\tau$  & $\rho_{s}$ & $\rho_{p}$ & $\tau$ & $\rho_{s}$ & $\rho_{p}$ & $\kappa$\\
\midrule
\textbf{CLIP-S} & \texttt{openai/clip-vit-large-patch14}  & 0.200 & 0.253 & 0.209 & 0.333 & 0.356 & 0.418 & -0.096 \\
\textbf{LLaVA-7b}  & \texttt{v1.5-7b}  & 0.400 & 0.532 & 0.633 & 0.483 & 0.597 & 0.507 & -0.003 \\
\textbf{LLaVA-13b}  & \texttt{v1.5-13b} & 0.433 & 0.559 & 0.659 & 0.517 & 0.618 & 0.523 &  0.010 \\
\textbf{GPT-4V} & \texttt{1106-vision-preview} & 0.400 & 0.544 & 0.540 & 0.500 & 0.594 & 0.470 & 0.080\\
\bottomrule
\end{tabular}
}
\end{table*}

\section{Experiments}
We have undertaken an empirical comparison between human assessors and vision-language models to offer an initial evaluation of their current capabilities in estimating relevance judgments. 
This comparative analysis encompasses one embedding-based model (CLIP) and two LLMs trained by visual instruction tuning (LLaVA and GPT-4V). 
The experiments were carried out in January 2024.

\subsection{Setups}

\paragraph{Test Collection.}
Our study focuses on the image suggestion task in TREC-AToMiC 2023.
In this task, queries are sections from Wikipedia pages, and the corpus contains images from Wikipedia. 
We assess VLMs' ability to assign relevance labels to 9,818 image--text pairs across 74 test topics.
We predict relevance scores, generate qrels for 16 retrieval runs, and compare them with NIST human-assigned qrels.
Note that the test topics consist of Wikipedia text sections (level-3 vital articles) without accompanying images, and NIST qrels are not publicly accessible during the training of VLMs we study in this work.

\paragraph{Vision--Language Models.}
Our experiments feature three models: CLIP~\cite{radford2021learning}, LLaVA~\cite{liu2023llava,liu2023improved}, and GPT-4V~\cite{gpt4v}. 
CLIP serves as a versatile baseline model, offering similarity scores for image--text pairs. 
We use CLIPScore~\cite{hessel-etal-2021-clipscore} (referred to as CLIP-S) for calculating relevance with CLIP. 
However, CLIP has limitations due to its text encoder's token limit (77 tokens), making it less adaptable for complex tasks with lengthy contexts. 
In contrast, LLMs like LLaVA and GPT-4V, fine-tuned for visual instruction understanding, possess larger text encoders capable of handling extended context. 
These models excel in various vision-language tasks, making them more versatile compared to CLIP.

\subsection{Correlation Study}
In this subsection, our primary aim is to investigate the extrinsic properties of relevance judgments generated by various approaches, where we base our analysis on retrieval runs and ranking metrics.
While various techniques exist to enhance the capabilities of vision-language models, including prompt engineering, few-shot instructions, and instruction tuning, our current focus centers on examining their zero-shot capabilities. 
We defer the exploration of other methods to future research endeavors. 
Following the work of~\citet{voorhees1998variations}, we undertake an investigation into the system ranking correlation and the agreement between the relevance labels estimated by the model and those provided by NIST annotators.
We evaluate the ranking correlations concerning the primary metrics utilized in the AToMiC track: NDCG@10 and MAP, and calculate Kendall's $\tau$, Spearman's $\rho_{s}$, and Pearson's $\rho_{p}$. In our agreement study, we compute Cohen's $\kappa$ using NIST's qrels as references. 

\paragraph{Overall.}
The primary results are showcased in Table~\ref{tab:main}, where rows correspond to the backbone model used for relevance judgment generation. 
Notably, models leveraging LLMs such as LLaVA and GPT-4V outperform the CLIP-S baseline concerning ranking correlation. 
Specifically, they achieve Kendall's $\tau$ values of approximately 0.4 for NDCG@10 and around 0.5 for MAP.
For comparison, previous research reported 0.9 for $\tau$ for MAP when comparing two types of human judgments~\cite{voorhees1998variations}. 
While there is still room for further improvement, our observations already demonstrate enhancement compared to the CLIP-S baseline: 0.200 (0.333) for NDCG@10 (MAP).
Moreover, other correlation coefficients, including Spearman and Pearson, corroborate the trends identified by Kendall's $\tau$. 
Additionally, we notice a rising trend in agreement levels when transitioning from CLIP-S (-0.096) to GPT-4V (0.080), as evidenced by Cohen's $\kappa$ values. 
The agreements achieved by the two largest models (LLaVA-13b and GPT-4V) are categorized as 'slight,' which represents an improvement over the smaller LLaVA-7b model and the baseline.

\begin{figure}[h!]
    \centering
    % \vspace{-0.2cm}
    \resizebox{\columnwidth}{!}{
    \includegraphics{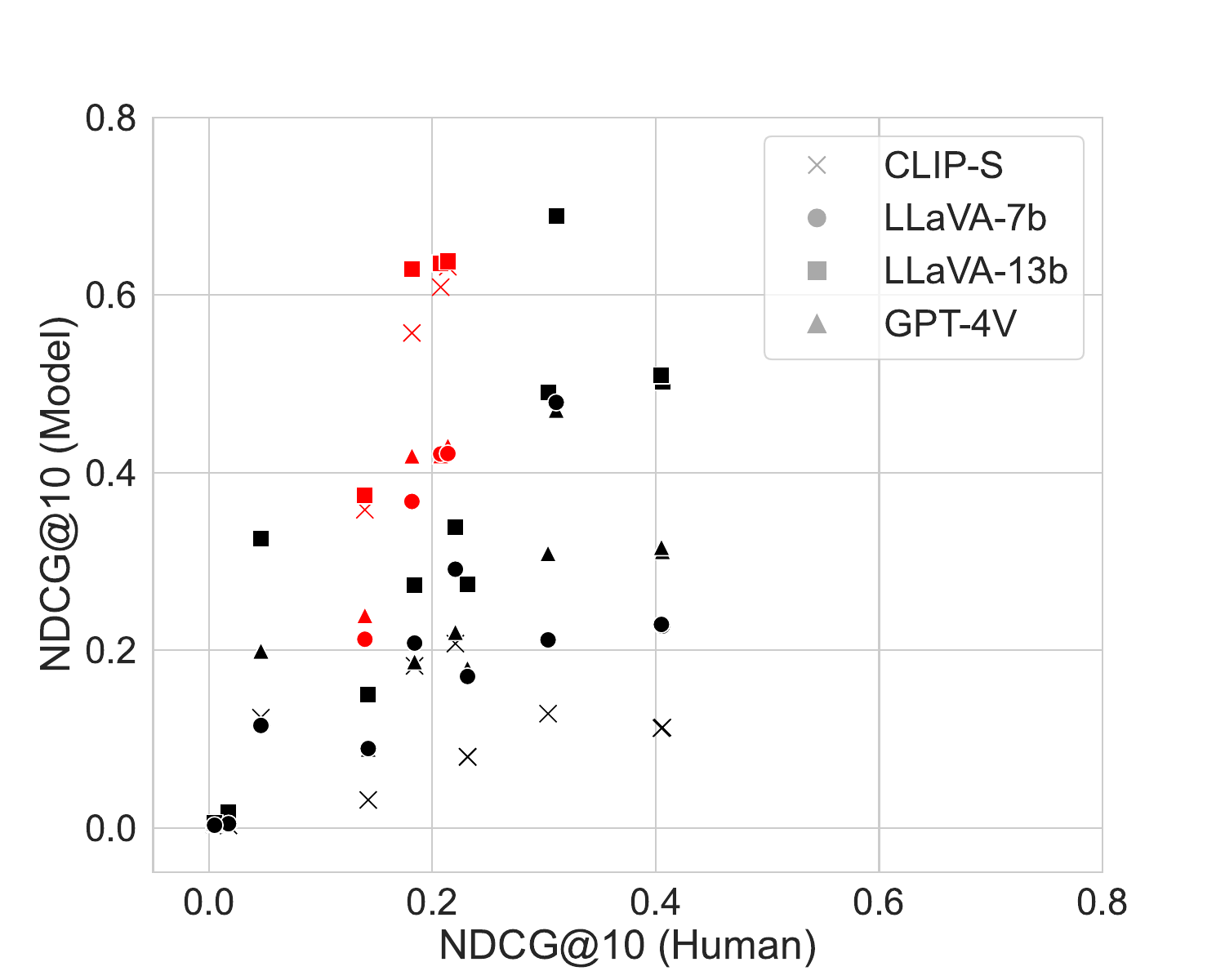}
    }
    \caption{Scatter plots of effectiveness (NDCG@10) for TREC-AToMiC 2023 runs using human-based and model-based qrels. Each data point represents the mean effectiveness of a single run evaluated with different qrels. CLIP-based runs are highlighted in red. Best viewed in color.}
    % \vspace{-0.2cm}
    \label{fig:scatter}
\end{figure}

\begin{table}[t]
    \centering
    % \small
    \caption{Evaluation bias assessment using Relative $\Delta$ in terms of NDCG@10 and MAP. A positive $\Delta$ favors CLIP-based systems, while a negative $\Delta$ favors other types of systems.}
    \resizebox{.7\columnwidth}{!}{
    \begin{tabular}{lrr}
    \toprule
    Model              &  $\Delta$(NDCG@10) & $\Delta$(MAP) \\
    \midrule
    \textbf{CLIP-S}    & \textbf{114.7} & \textbf{120.5} \\
    \textbf{LLaVA-7b}  & 58.5 & 86.6 \\
    \textbf{LLaVA-13b} & 55.8 & 83.1 \\
    \textbf{GPT-4V}    & 64.0 & 91.3 \\
    \textbf{Human}     & -11.7 & -19.5 \\
    \bottomrule
    \end{tabular}
    }
    \label{tab:bias}
\end{table}

\paragraph{Evaluation Bias}
Model-based evaluations can introduce bias, often favoring models that are closely related to the assessor model~\cite{liu2023gpteval,pangakis2023automated}.
We term this phenomenon as \textit{evaluation bias}. 
This is distinct from \textit{source bias} which indicates that neural retrievers might prefer contents generated by generative models~\cite{dai2023llms}.
To address this potential concern, we conducted an initial analysis using the scatter plot presented in Fig.~\ref{fig:scatter}. 
In this analysis, we compared the NDCG@10 scores of the 16 submissions made by participants employing different sets of qrels. 
Each data point on the plot corresponds to a specific run, with distinct markers representing variations in results based on relevance estimation models.
Upon closer examination of the plot, we identified a positive correlation between model-based and human-based qrels. 
Notably, the effectiveness of submitted systems appeared slightly higher when compared to those using human-based qrels. 

To gain deeper insights, we've visually highlighted CLIP-based submissions in red for a thorough investigation. 
This visual distinction underscores the preference for model-based qrels for CLIP-based systems, especially evident with CLIP-S qrels.
We quantitatively assess this bias using a metric adapted from the work of~\citet{dai2023llms}:

\begin{equation*}
    \resizebox{\hsize}{!}{%
    $\text{Relative}\ \Delta = 2 \frac{\text{Metric}_\text{CLIP-based} - \text{Metric}_\text{Others}}{\text{Metric}_\text{CLIP-based} + \text{Metric}_\text{Others}} \times 100\%$,%
}
\end{equation*}
here Metric stands for a measure, e.g., NDCG@$k$, averaged across systems.
Observing Table~\ref{tab:bias}, CLIP-S exhibits a strong bias, with Relative $\Delta = 114.7$ for NDCG@10 and 120.5 for MAP.
LLM-based approaches also display a slight bias towards CLIP-based systems, possibly because both LLaVA and GPT-4V rely on CLIP embeddings for image representations.
In contrast, human-based qrels show the lowest bias: -11.7 for NDCG@10 and -19.5 for MAP.

\begin{figure}[h]
    \centering
    % \vspace{-0.2cm}
    \resizebox{\columnwidth}{!}{
    \includegraphics{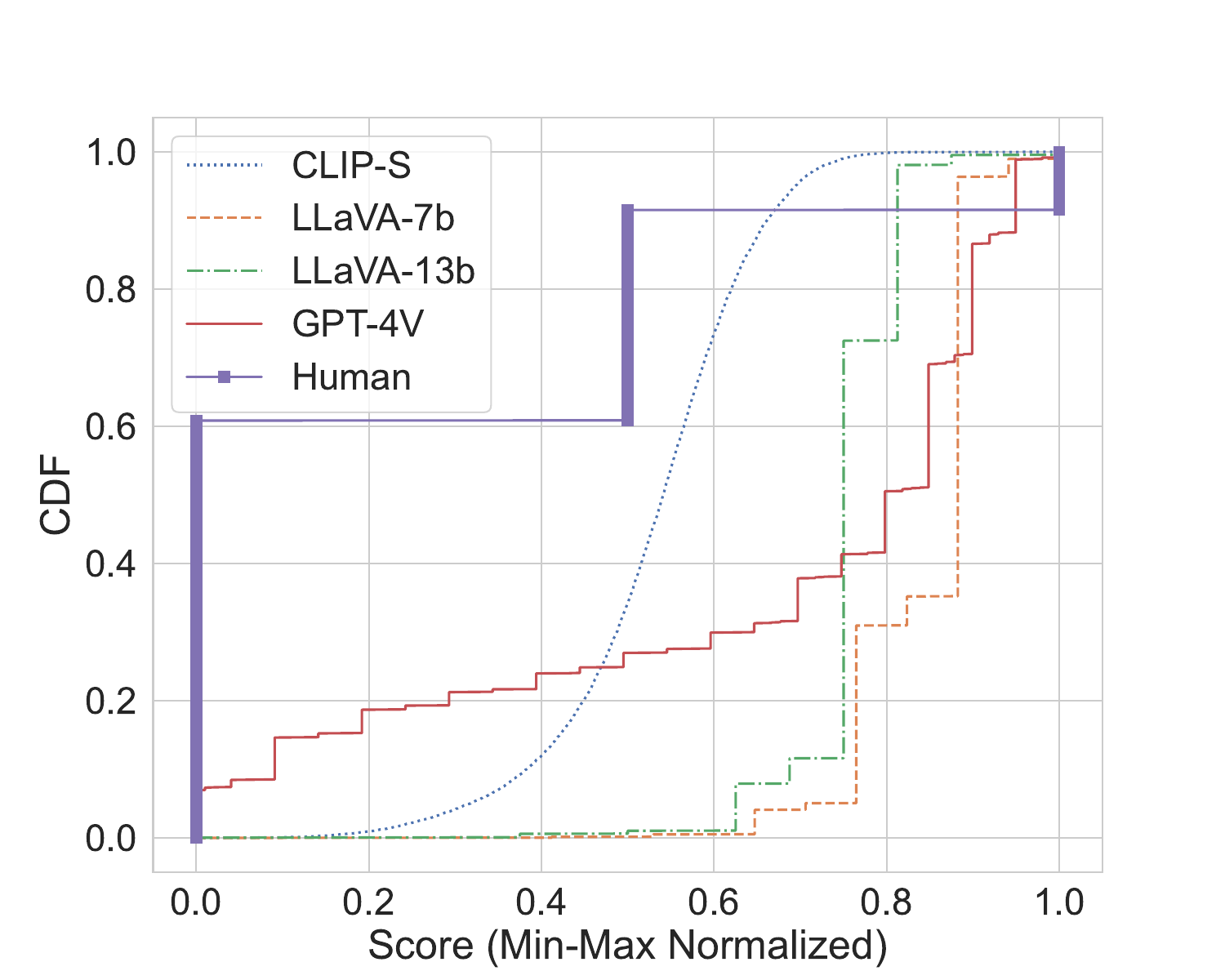}
    }
    \caption{Cumulative distribution function (CDF) plot of relevance scores from various models. Human stands for relevance annotations of NIST qrels.}
    \label{fig:cdf}
    % \vspace{-0.2cm}
\end{figure}

\begin{figure}[ht]
    \centering
    \resizebox{\columnwidth}{!}{
    \includegraphics{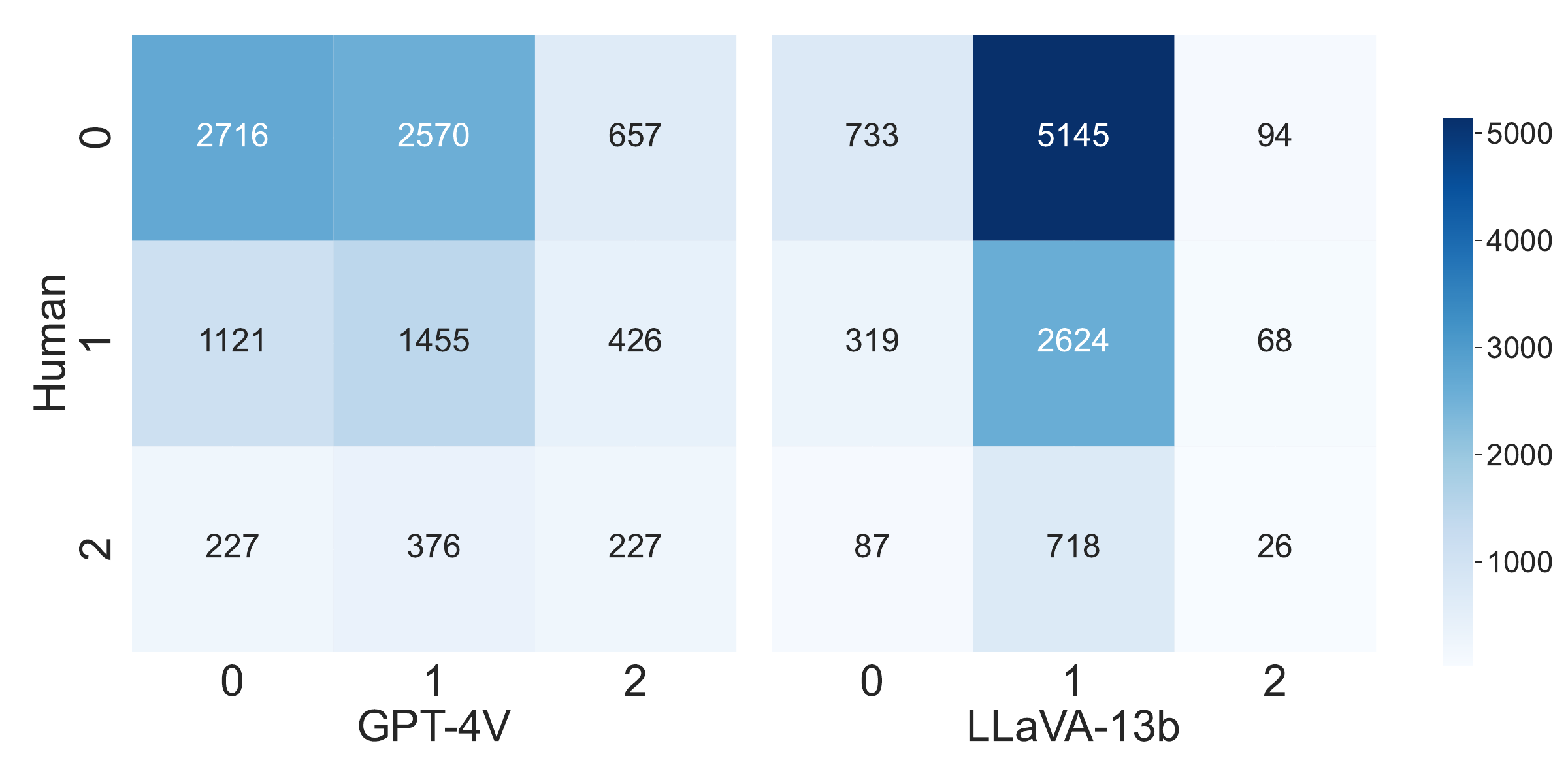}
    }
    \caption{Confusion matrices comparing human-based and model-based qrels. Tick labels 0/1/2 represent Non-relevant/Related/Relevant graded levels. Best viewed in color.}
    \label{fig:cm}
    % \vspace{-0.2cm}
\end{figure}

\subsection{Estimated Relevance Analysis}
In this subsection, we aim to explore the intrinsic properties of relevance judgments generated by various systems. We began our analysis by examining score distributions, visualized in Figures~\ref{fig:cdf} and~\ref{fig:cm}, to gain insights into model-based scores.

Figure~\ref{fig:cdf} presents a Cumulative Distribution Function (CDF) plot of scores before post-processing into relevance levels (0, 1, and 2). We included NIST qrels (human) results for reference. 
Notably, GPT-4V's score distribution closely aligns with the human CDF, while CLIP-S exhibits a smoother S-shaped distribution with limited representation of low-relevance data. 
LLaVA produces tightly concentrated scores, adding complexity to post-processing, particularly when compared to GPT-4V.

Figure~\ref{fig:cm} illustrates confusion matrices, highlighting LLaVA's tendency to generate more 1 (related) judgments, fewer 2 (relevant), and 0 (non-relevant) judgments compared to GPT-4V. 
We anticipate that future models will strive to produce score distributions that better match human annotations, thereby addressing these challenges and limitations. 
Further studies~\cite{zhuang2023beyond} on harnessing LLMs' relevance prediction capability are necessary.

\section{Conclusion}
This study delves into the capabilities of VLMs such as CLIP, LLaVA, and GPT-4V for automating relevance judgments in image--text retrieval evaluation. 
Our findings reveal that visual-instruction-tuned LLMs outperform traditional metrics like CLIPScore in aligning with human judgments, with GPT-4V showing promise due to its closer alignment with human judgment distributions.

Despite these advancements and low cost of model-based relevance annotation,\footnote{The cost of using GPT-4V API for the experiments is around USD 150.} challenges such as evaluation bias and the complexity of mimicking human judgments remain. 
These issues underscore the need for ongoing model refinement and exploration of new techniques to enhance the reliability and scalability of automated relevance judgments.

In conclusion, our research highlights the potential of VLMs in streamlining multimedia content creation while also pointing to the critical areas requiring further investigation. 
The path toward fully automated relevance judgment is complex, necessitating continued collaborative efforts in the research community to harness the full potential of VLMs in this domain.

\section*{Acknowledgements}
This research was supported in part by the Canada First Research Excellence Fund and the Natural Sciences and Engineering Research Council (NSERC) of Canada.

% \balance

% \bibliographystyle{acm}
\bibliography{ref}

\end{document}